\documentclass[useAMS,usenatbib]{mn2e}
\usepackage{graphicx}
\usepackage{float}
\usepackage{footnote}

\title[The Closest Look at 1H0707-495: X-ray Reverberation Lags with 1.3 Ms of Data]{The Closest Look at 1H0707-495: \\X-ray Reverberation Lags with 1.3 Ms of Data}
\author[Kara et al.]{E. Kara$^{1}$\thanks{E-mail:
ekara@ast.cam.ac.uk}, A. C. Fabian$^{1}$, E. M. Cackett$^{2}$, J. F. Steiner$^{1}$, P. Uttley$^{3}$,\newauthor{D. R. Wilkins$^{1}$ and A. Zoghbi$^{4,5}$}\\
$^{1}$Institute of Astronomy, Madingley Rd, Cambridge CB3 0HA\\
$^{2}$Department of Physics and Astronomy, Wayne State University, Detroit, MI 48201, USA\\
$^{3}$Astronmical Institute `Anton Pannekoek', University of Amsterdam, Postbus 94249, 1090 GE Amsterdam, the Netherlands\\
$^{4}$Department of Astronomy, University of Maryland, College Park, MD 20742, USA\\
$^{5}$Joint Space-Science Institute (JSI), College Park, MD 20742-2421, USA}

\begin{document}

\date{Accepted 2012 October 4.  Received 2012 October 4; in original form 2012 July 26}

\pagerange{\pageref{firstpage}--\pageref{lastpage}} \pubyear{2012}

\maketitle

\label{firstpage}

\begin{abstract}
 Reverberation lags in AGN were first discovered in the NLS1 galaxy, 1H0707-495. We present a follow-up analysis using 1.3 Ms of data, which allows for the closest ever look at the reverberation signature of this remarkable source.  We confirm previous findings of a hard lag of $\sim~100$ seconds at frequencies $\nu \sim~[0.5 - 4] \times 10^{-4}$~Hz, and a soft lag of $\sim~30$ seconds at higher frequencies, $\nu \sim~[0.6 - 3] \times 10^{-3}$~Hz. These two frequency domains clearly show different energy dependences in their lag spectra.  We also find evidence for a signature from the broad Fe K$\alpha$ line in the high frequency lag spectrum.  We use Monte Carlo simulations to show how the lag and coherence measurements respond to the addition of Poisson noise and to dilution by other components.  With our better understanding of these effects on the lag, we show that the lag-energy spectra can be modelled with a scenario in which low frequency hard lags are produced by a compact corona responding to accretion rate fluctuations 
propagating through an optically thick accretion disc, and the high frequency soft lags are produced by short light-travel delay associated with reflection of coronal power-law photons off the disc.

\end{abstract}

\begin{keywords}
black hole physics -- galaxies: active -- X-rays: galaxies -- galaxy: individual : 1H0707-495.
\end{keywords}

\section{Introduction}

Recent X-ray variability studies of active galactic nuclei (AGN) have given us unique insights into the energetics and geometry of the innermost regions of accreting black hole systems. By measuring the phase lags between light curves at different energy bands, we can probe distances as small as tens of lightseconds in AGN that are billions of lightyears away. Narrow-line Seyfert I galaxy 1H0707-495 ($z = 0.0411$) was the first AGN shown to exhibit reverberation lags \citep{fabian09}, and since then, this feature has been significantly detected in 15 other Seyfert galaxies \citep{emmanoulopoulos11,demarco11,zoghbi11b,demarco12,zoghbi12}. Further evidence for reverberation lags of milliseconds or less have been seen in stellar mass black hole X-ray binary GX~339-4 \citep{uttley11}.  
 
Reverberation lags can be explained by the standard reflection model.  Matter in a geometrically thin, optically thick accretion disc is accreted onto the black hole, causing the accretion disc to heat up and radiate, mostly in the ultraviolet \citep{shakura73}.  Those thermal disc photons are Compton upscattered to X-ray energies in the corona or base of the jet, producing the observed X-ray power-law continuum. Some continuum photons are directed back towards the accretion disc, causing it to fluoresce with the characteristic reflection spectrum. Soft lags or `reverberation lags', where the power-law continuum leads the `soft excess' associated with the photoionized disc reflector, are thought to be due to the effect of the light travel time between the corona and the inner accretion disc. Hard lags, where the power-law continuum follows variations at soft energies, have been explained as the propagation of mass accretion rate fluctuations, which modulate regions emitting soft photons before those emitting hard 
photons \citep{kotov01,arevalo06}. 

{\em XMM-Newton} observations of 1H0707-495 have been some of the most important resources in gaining an understanding of the X-ray emission from the innermost regions of active galaxies. The first {\em XMM-Newton} observations of 1H0707-495 showed a sharp drop in the spectrum at $\sim~7$~keV \citep{boller02}. Two interpretations were developed to explain this feature, namely an absorption edge from a partial covering absorber \citep{tanaka04}, and the edge to the blue wing of a broad iron K$\alpha$ line, produced in the inner regions of the disc \citep{fabian04}. 

Further observations in 2008 showed more evidence for the inner disc interpretation, as the complementary broad iron L$\alpha$ line was also clearly identified \citep[][]{fabian09,zoghbi10}.  In addition to this spectral evidence, timing studies showed a $\sim30$ second light travel delay between the power law source and the reflection from the disc. \citet{miller10} interpreted this time delay as being due to scattering in a more distant reflector/absorber associated with a wind directed along the line of sight, though the subsequent X-ray reverberation detections in several different AGN make this an unlikely scenario.  See \citet{zoghbi11} for further discussion regarding problems with this model.

In 2011, 1H0707-495 went into a low flux state, which triggered an observation with {\em XMM-Newton}.  The spectral drop associated with the blue wing of the  Fe K$\alpha$ appeared at $\sim~6$~keV, suggesting emission very close to the central black hole, within 1 $r_{\mathrm{g}}$ of the event horizon \citep{fabian12}. Lastly, recently public observations from 2010 confirm reflection from a highly ionized disc, and suggest that the reflection spectrum could be slightly modified by a highly ionized outflow \citep{dauser12}.

In this paper, we compile all the {\em XMM-Newton} observations from 2000 to 2010 ($> 1.3$~Ms), to take the closest ever look at the reverberation lags in 1H0707-495. With $\sim$~2.5 times the dataset, we are better able to define the lag spectrum and structures in the data that were suggested by the analysis of \citet{zoghbi11}.  The observations and data reduction are described in Section~\ref{observations}, and the results are presented in Section~\ref{results}. We continue with a study of the robustness of the lag measurement using Monte Carlo simulations in Section \ref{simulations}, and discuss our interpretation through a physical model for the lag spectrum in Section \ref{lag_model}.

\section[]{Observations and Data Reduction}
\label{observations}

1H0707-495 has been observed several times over the past 12 years with the {\em XMM-Newton} satellite \citep{jansen01}. The 14 observations used in this analysis are shown in Table~\ref{obs}. 
We focus on the data from the EPIC-PN camera \citep{struder01}, which were all made in large window imaging mode, with the exception of the 2000 and 2001 data, which were taken in full window imaging mode.  All the data were reduced in the same way using the {\em XMM-Newton} Science Analysis System (SAS v.11.0.0) and the newest calibration files.
The data were cleaned for high background flares,  and were selected using the condition {\sc pattern} $\le 4$. No significant pile-up was found in any of the observations.

\begin{table}
\centering
\begin{tabular}{c|c|c}
\hline
Obs. ID & Date (UT) & Exposure (s)\\
\hline
0110890201 & 2000-10-21 00:38:28 & 46018 \\ 
0148010301 & 2002-10-13 11:10:37 & 79953 \\
0506200301 & 2007-05-14 02:19:28 & 40953 \\
0506200201 & 2007-05-16 06:08:30 & 40914 \\
0506200501 & 2007-06-20 13:11:46 & 46913 \\
0506200401 & 2007-07-07 10:40:03 & 42866 \\
0511580101 & 2008-01-29 18:28:24 & 123815 \\
0511580201 & 2008-01-31 16:29:46 & 123670 \\
0511580301 & 2008-02-02 18:22:11 & 122504 \\
0511580401 & 2008-02-04 18:24:21 & 121922 \\
0653510301 & 2010-09-13 00:32:10 & 116575 \\
0653510401 & 2010-09-15 00:34:12 & 128200 \\
0653510501 & 2010-09-17 00:36:42 & 127602 \\
0653510601 & 2010-09-19 00:35:36 & 129001 \\
\hline
\end{tabular}
\caption{The 14 {\em XMM-Newton} observations used in this analysis, amounting to 1.3~Ms of data.}
\label{obs}
\end{table}

The source spectra were extracted from circular regions of radius 35 arcsec centered on the maximum source emission, and the background spectra were chosen from background regions that were the same distance to the readout node as the source region.  In order to get the best description of the background, we chose the background regions to be as large as possible, often twice the radius of the source region.  The position of the background regions were chosen to avoid the Cu-K emission lines from the electronic circuits behind the PN CCD that contaminate the background at 8.0 and 8.9~keV.  Despite these efforts, the Cu-K line still mildly contaminates the data, and so we take heed with regards to the highest energy bin at 7-10 keV. 

The response matrices were produced using {\sc rmfgen} and {\sc arfgen} in {\sc SAS}.  The spectra from all observations were merged before fitting using {\sc mathpha} in {\sc FTOOLS}, and the resulting combined spectra were rebinned to contain a minimum of 20 counts per bin.  Because the observations differ significantly in flux, we also used {\sc tabgtigen} in {\sc SAS} to filter the observations by rate, in order to check the flux resolved spectra and compare them to the combined spectrum from the 14 observations.  Spectral fitting was performed using {\sc xspec} v12.5.0 \citep{arnaud96}. All quoted errors correspond to a 90\% confidence level, and energies are given in the rest frame of the source. The quoted abundances refer to solar abundances in \citet{anders89}.

\section{Results}
\label{results}

\subsection{Lag vs. Frequency Spectrum}
\label{freq}

\begin{figure}
\includegraphics[width=\columnwidth]{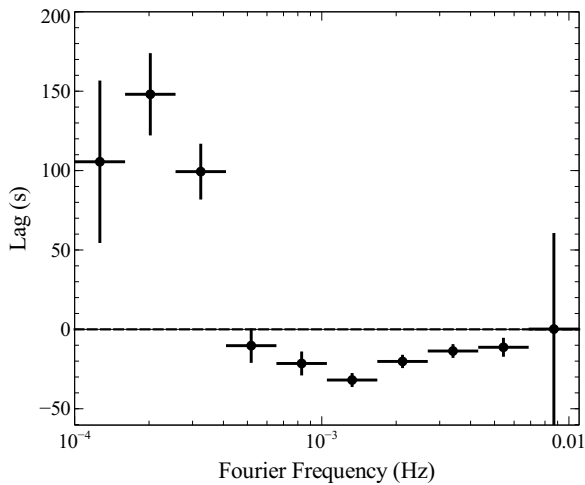}
\caption{Lag-frequency spectrum using 1.3~Ms of data. The lag is calculated between the soft energy band (0.3 - 1.~keV) and the hard band (1.2 - 4.~keV). We adopt the convention that negative lag mean the soft band lags behind the hard band. The most negative lag (at $1.33 \times 10^{-3}$~Hz) is $-31.9 \pm 4.2$ s. }
\label{lag_freq}
\end{figure}

Using the light curves of the 14 observations, we compute the Fourier phase lag between the hard and soft energy bands, following the technique described in \citet{nowak99}.  
We write the Fourier transform of the soft light curve, $s$, as $S=| S | e^{i\phi_s}$. The product of $S$ with the complex conjugate of the hard band FFT, $H^{\ast}$, is written, $ H^{\ast} S = | H | | S | e^{i(\phi_s-\phi_h)}$, which gives us the phase difference between the two signals.  The overall Fourier phase lag, $\phi(f)$ is the phase of the average cross power spectrum. That is, $\phi(f) = \mathrm{arg}[\langle H^{\ast}(f)S(f) \rangle]$. We then convert this to a time lag, $\tau \equiv \phi(f)/2\pi f$. We note that the Fourier phase is defined on the interval $(-\pi,\pi)$, which can cause phase wrapping at high frequencies \citep{nowak96}.
 
Fig.~\ref{lag_freq} shows the lag-frequency spectrum between the energy bands 0.3 -- 1.0~keV (soft-excess dominated) and 1.2 -- 4.0~keV (power law dominated).  The light curve segments range in individual exposure times from 40000 seconds to 110000 seconds with 10 second bins. The frequency range goes up to 0.01~Hz, beyond which the signal is dominated by Poisson noise. As seen in previous timing studies of 1H0707-495 by \citet{fabian09}, \citet{zoghbi10} and \citet{emmanoulopoulos11}, the hard flux is shown to lag behind the soft by hundreds of seconds at frequencies less than $\sim~5 \times 10^{-4}$~Hz (hereafter referred to as the ``hard lag'').  Above frequencies of $\sim~7 \times 10^{-4}$~Hz, the soft flux is clearly shown to lag the hard by roughly 30 seconds (referred to as the ``soft lag'').

\subsection{Lag vs. Energy Spectrum}

\begin{figure}
\begin{center}
\includegraphics[width=\columnwidth]{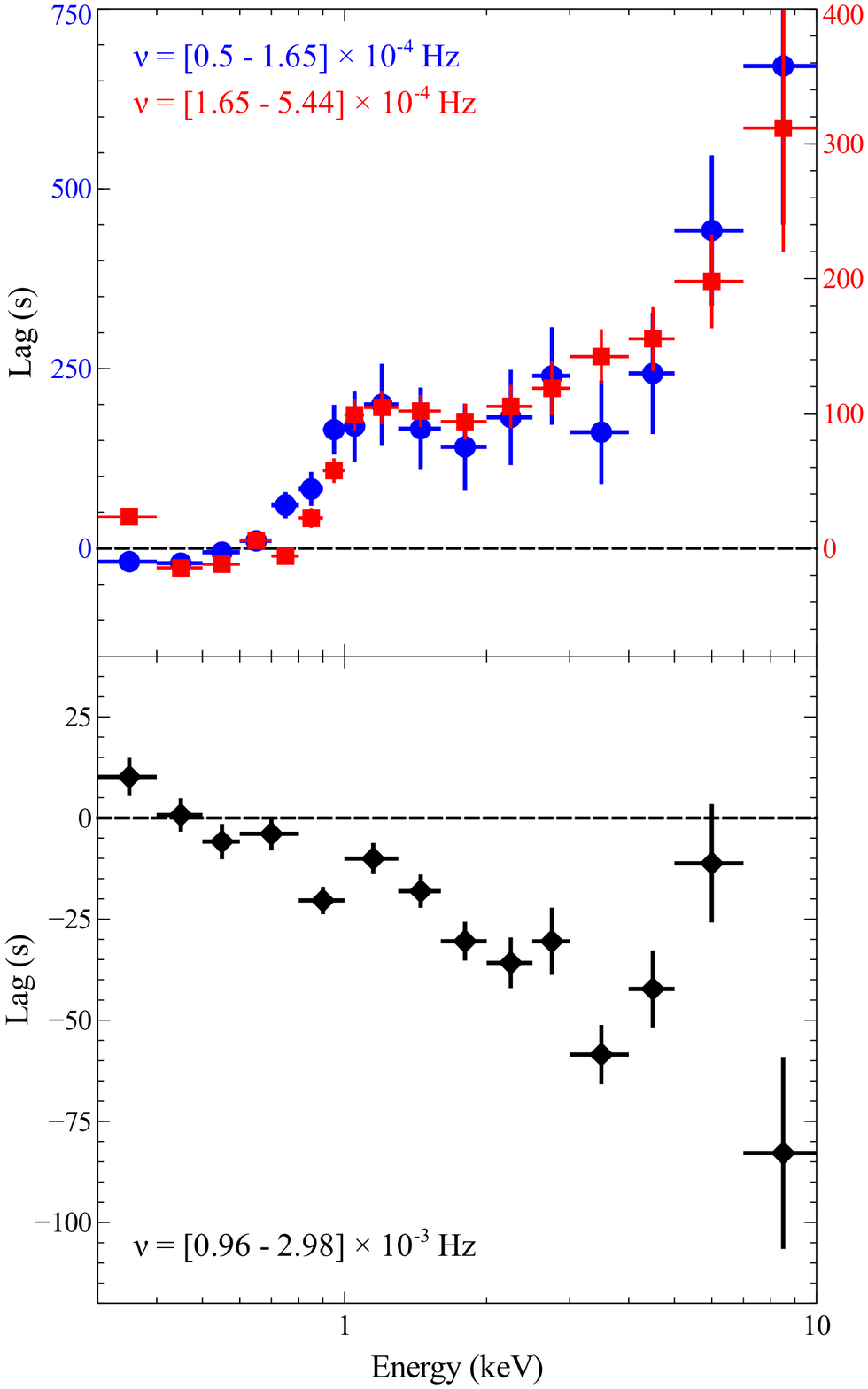}
\caption{
Lag-energy spectra showing the energy dependence of the lags at particular frequencies. 
(a) The lag-energy spectrum for very-low frequencies, $\nu = [0.5 - 1.65] \times 10^{-4}$~Hz (blue circle; left axis), and low frequencies, $\nu = [1.65-5.44] \times 10^{-4}$~Hz (red square; right axis). The upward trend means that the low energy component leads the high energies at these long timescales (b) high frequency lag-energy spectrum for $\nu = [0.96 - 2.98] \times 10^{-3}$~Hz.  The nearly zero-lag at low energies, and again at $\sim~6.5$~keV mean that the signal at these energies respond at the same time as the reference band.}  

\label{lag_spec}
\end{center}
\end{figure}

Taking a closer look, we examine how the time lag evolves with energy, using roughly equal logarithmically-spaced energy bands (henceforth `energy bins') in a particular frequency range, similar to \citet{zoghbi11}.  For each energy bin, the lag is computed between the light curve in that bin and the light curve of a broader reference band from 0.3 -- 0.8~keV. This reference energy range is selected because it is the band with the highest signal-to-noise. For lags computed in energy bins between 0.3 -- 0.8~keV, the light curve of the energy bin is removed from the reference to ensure that the Poisson noise remains uncorrelated. The lags have not been shifted, and therefore zero-lag means that there is no time delay between that bin and the soft excess reference band. Similarly, a negative lag means that that bin leads the reference, and positive lags mean that that bin lags behind the reference.  Previous lag-energy studies by \citet{zoghbi11} use the entire energy range as the reference band. We find that this does not change 
the shape of the lag-energy profile, but does make for an arbitrary constant lag shift in the entire spectrum.  By choosing the reference band from 0.3 -- 0.8~keV, we have a more physical interpretation of what the value of the lag means, namely that it lags or leads the soft excess.  

The top panel of Fig.~\ref{lag_spec} shows the energy dependence of the lag for the low frequencies where we see the hard lag.  The lag-energy spectrum is shown in two frequency ranges: for the `low frequency range' ($\nu = [1.65.-5.44] \times 10^{-4}$~Hz), and `very-low frequency range' ($\nu = [0.5 - 1.65] \times 10^{-4}$~Hz). The lag spectrum exhibits a general positive trend with increasing energy. We see a nearly constant zero-lag from bins in the 0.3 -- 0.9~keV range.
At 1~keV, the lag jumps up hundreds of seconds, meaning that the signal from 1 -- 2~keV is occurring hundreds of seconds after the reference band.  The lag continues to increase with energy up to 10~keV.

The low frequency and the very-low frequency lag spectra have the same general shape, but the lag timescales are twice as long for the very-low frequency range.  For the case of propagating viscous fluctuations, this corresponds to variability on long timescales occurring further away from the centre. We also notice that the profile differs at low energies between the two frequency ranges. The low frequency lag shows a clear step up to $\sim~100$ seconds, where the very-low frequency lag shows a more gradual increase up to the constant lag at 1~keV.  This suggests that individual components can contribute more at specific frequencies.  In this case, the soft excess (due to thermal or reflected emission) dominates more at the low frequencies, closer to the central region, 
than the very-low frequencies, from further out. 

The bottom panel of Fig.~\ref{lag_spec} shows the energy dependence of the lag at higher frequencies from $[0.96 - 2.98] \times 10^{-3}$~Hz (i.e. the soft lag frequencies). This lag-spectrum has a general negative trend, starting at around zero lag in the bins from 0.3 -- 0.7~keV, and decreasing to $\sim-60$ seconds in the 3 -- 4~keV bin, after which it peaks to zero-lag at the energy of the Fe K$\alpha$ line at $\sim~6$~keV.  Simulations of lags between coherent light curves with the same noise level to that of the 7 -- 10 keV bin suggest despite low signal-to-noise, the lag can be recovered.  Though we have taken the steps to remove contamination by the Cu-K line of the electronics in the 7 -- 10 keV bin, some residuals might effect measurements in this bin.

It is important to note that the high-frequency lag spectrum looks very different than the one at lower frequencies, suggesting that different physical processes contribute to the hard versus soft lags.

The high and low frequency lag spectra found here with the 1.3~Ms sample are similar to those in \citet{zoghbi11}, but are much more clearly defined.  Of special interest is the high-frequency feature at $\sim$~6~keV, the energy of the iron K$\alpha$ line. A clear lag signature from the Fe L line is not seen in the data, and this is likely because of the thermal reprocessed emission prominent at low energies \citep[][ and see Section~\ref{lag_model}]{zoghbi11}.

\subsection{Covariance Spectrum}
\label{covariance}

\begin{figure}
\includegraphics[width=\columnwidth]{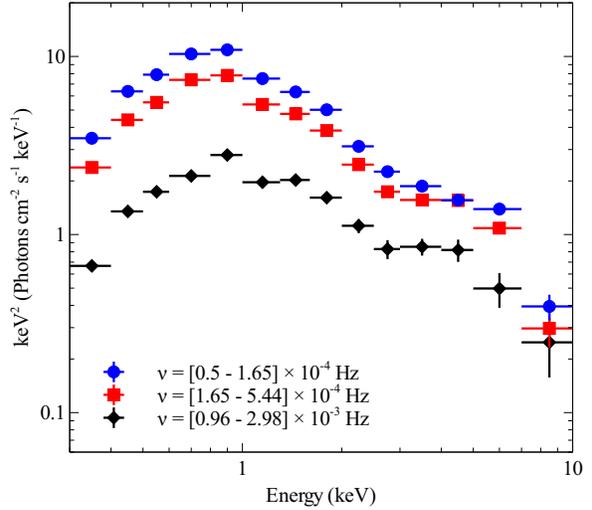}
\caption{Covariance spectrum for very-low frequencies $\nu = [0.5 - 1.65] \times 10^{-4}$~Hz (blue circle), low frequencies $\nu = [1.65 - 5.44] \times 10^{-4}$~Hz (red square), and high frequencies $\nu = [0.96 - 2.98] \times 10^{-3}$~Hz (black diamonds). At low energies the error bars are much smaller than the symbols. 
We fit a simple power law model at 1 -- 3~keV. For very-low and low frequencies $\Gamma \sim~3.2$, and for high frequencies $\Gamma \sim~2.9$.}
\label{covspec}
\end{figure}

\begin{table}
\centering
\begin{tabular}{c|c}
\hline
Frequency (Hz) & Photon Index, $\Gamma$\\
\hline
$\nu = [0.5 - 1.65] \times 10^{-4}$ & $3.23 \pm 0.04$ \\
$\nu = [1.65 - 5.44] \times 10^{-4}$ & $3.12 \pm 0.05$ \\
$\nu = [0.96 - 2.98] \times 10^{-3}$ & $2.89 \pm 0.16$ \\
time-integrated & $3.228 \pm 0.009$ \\
\hline
\end{tabular}
\caption{Power law photon indices for the covariance spectra in three frequency ranges, and the time-integrated spectrum for comparison.}
\label{tab:cov_pl}
\end{table}

The covariance spectrum is a measure of the absolute amplitude of correlated variations in count rate as a function of energy \citep[see][]{wilkinson09,uttley11}. Using this approach, we can examine the time-varying portion of the energy spectrum that is correlated with the reference band at a particular frequency range.  Note that the covariance spectrum is not simply measuring the fractional variability amplitude, as in an rms spectrum, but rather is proportional to the photon count rate, allowing us to make a direct comparison with the time-integrated spectrum.

Fig.~\ref{covspec} shows the covariance spectrum at the two low frequency ranges and the high frequency range.
We fit a simple power law model at 1 -- 3~keV for each spectrum in order to measure the power-law continuum without any contamination from the red wing of the Fe K$\alpha$ line. The photon indices are shown in Table \ref{tab:cov_pl}.  We found that the  photon index hardens with increasing frequency from $\Gamma = 3.23$ to $\Gamma = 2.89$.  
We will discuss further implications of this result in Section~\ref{lag_model}. The same spectral hardening is suggested when fitting the data from 1.3 -- 3 keV alone, though the change is not significant.

Of the three frequency ranges shown here, the high-frequency covariance spectrum shows the most prominent peak at 0.9~keV, the energy of the Fe~L line. The spectrum also appears to increase again at 3 -- 4~keV, the energy of the red wing of the Fe K$\alpha$ line. These observations further hint that the high frequencies are probing the inner accretion disc, where relativistic reflection is greatest.

\subsection{Coherence}

\begin{figure}
\includegraphics[width=\columnwidth]{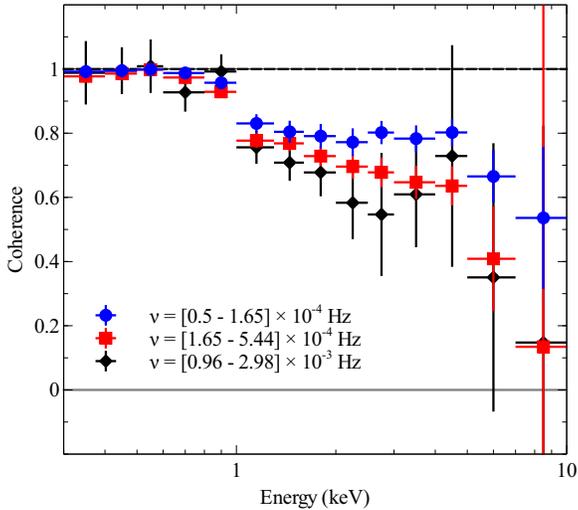}
\caption{ The coherence in each energy bin at very-low frequencies $\nu = [0.5 - 1.65] \times 10^{-4}$~Hz in blue circles, low frequencies $\nu = [1.65 - 5.44] \times 10^{-4}$~Hz in red squares, and high frequencies $\nu = [0.96 - 2.98] \times 10^{-3}$~Hz in black diamonds. 
In Section \ref{simulations}, we show that the lag can still be accurately recovered despite low coherence.}

\label{coher_spec}
\end{figure}

Coherence is a measure of the fraction of the variance of one process at a given frequency that is a linear transform of the other. In other words, a unity coherence shows that one time series can completely predict the other. In order to measure the lag between two signals, they must have some degree of coherence.  We calculate the coherence and error bars using Equation 8 from \citet{vaughan97}. Earlier studies by \citet{zoghbi10} examined the coherence as a function of frequency between the hard (1 -- 4~keV) and soft (0.3 -- 1~keV) band, and found that for these energies, the coherence is high, above $\sim~0.75$, up to $\sim~5 \times 10^{-3}$~Hz, where it drops sharply due to dominating Poisson noise.

Fig.~\ref{coher_spec} shows the coherence in each energy bin relative to the reference band (0.3 -- 0.8~keV). If the reference band is chosen to be the entire energy range (0.3 -- 10~keV), the shape of the coherence spectrum does not change, but the amplitude of the coherence become slightly higher.  This shows that any structure seen in the coherence spectra is not an artefact of the chosen reference band.

For all of the examined frequency ranges, the coherence is $\sim~1$ below 1~keV, and jumps down at 1~keV, suggesting that we are seeing contributions from distinct variable components.  

In the highest energy bins, the coherence drops, especially for high frequencies.  In the following sections, we describe our Monte Carlo simulations, which show that dilutions from other uncorrelated components can decrease the coherence, but the lag is still recovered. We therefore conclude that while some degree of coherence is necessary, unity coherence is not a strict requirement for an accurate measurement of the lag.

\section{Simulations}
\label{simulations}

We conducted simulations to understand the nuances of the lag calculation, and to gain clearer insights into the possible reasons for the structure seen in the lag spectra. In this section, we examine the robustness of the lag measurement using Monte Carlo simulations.

Simulated light curves were made from red-noise power spectra (PSD) using the \citet{timmer95} method, which allows for randomness in both the phase and amplitude of the Fourier transform. 
The underlying red-noise PSD has a spectral index of 1.6, consistent with the PSD of the data from 0.3 -- 10 keV.
In order to account for possible red-noise leakage into higher frequencies, the light curves were made $\sim~10$ times too long, and cut down to the average light curve length of the observations \citep{uttley02}.
The light curves were then scaled to the average count rate and variance of all the observations.
To simulate a lag, light curve pairs were generated from the same random realization of the Fourier transform.  A frequency-independent time delay was then added to the Fourier transform for one light curve. 
The lag and coherence was measured in the high frequency range $\nu = [0.96 - 2.98] \times 10^{-3}$~Hz. 

With these simulated light curves, we test the effects of Poisson noise and dilution on the measurements of coherence and lag.

\subsection{The effect of Poisson noise}

In order to understand the effect of Poisson noise on the measurement of the lag, we simulated 1000 coherent light curves pairs that were subjected to increasing amounts of noise. A constant time lag of 50 seconds was added to one of the light curves in the pair. 
We calculate the lag for each of the 1000 light curve pairs made in the simulation.  The mean lag was determined from the 1000 measured lags, and the standard deviation of the lag distribution calculated.  
Regardless of the Poisson noise, the mean lag recovered remains consistent with the input, though increasing the noise significantly increases the standard deviation of the recovered lags.
We test the recovery of the lag errors for 14 simulated light curve pairs (i.e. the same number of observations we use to calculate the lag), and find that the sizes of these error bars are consistent with the data.

\begin{figure*}
\includegraphics[width=\textwidth]{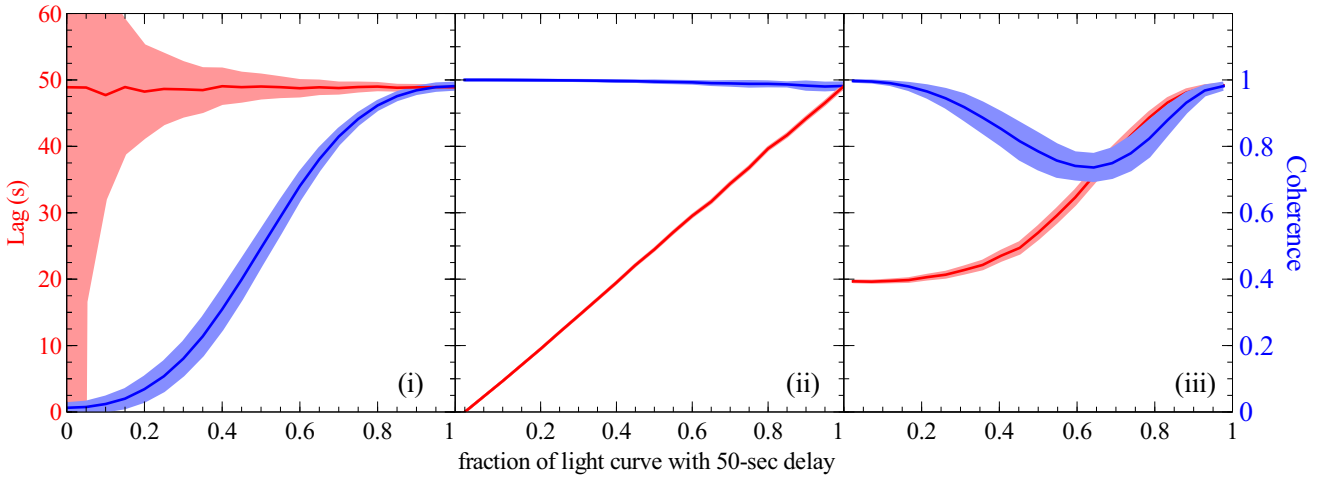}
\caption{ The recovery of the lag and coherence for diluted signals. When the fraction on the x-axis is 1, the signal is completely undiluted, and the 50 second lag is recovered with unity coherence. (i) The light curve with a 50 second delay is contaminated by a fraction of uncorrelated signal.  
(ii) The lag is measured between light curve pairs with 50 second offset that are being mixed together.  The recovered lag is a simple average of the lags of the different spectral components.  
(iii) The lag is computed between two variable, but uncorrelated light curve pairs, one pair is offset by 50 seconds, and the other is offset by 20 seconds.  
The coherence decreases with more mixing.  Shaded regions represent the standard deviation of the 1000 simulated lags.}
\label{dilution}
\end{figure*}

Poisson noise has the effect of adding a component with a random phase and amplitude to the Fourier transform of our signal. Since the lag is a measurement of the phase lag between coherent signals, an incoherent random phase does not effect the value of the lag. 

These simulations show that Poisson noise does not effect the value of the measured lag as a consistent estimator of the intrinsic lag. We find that at high frequencies and at high energies where there is low signal-to-noise, it is still possible to measure a meaningful lag, as long as appropriate error calculations are made.

\subsection{The effect of dilution} 
\label{dilution_sec}

Dilution is another possible effect on the measurement of the lag and coherence. If the signals are contaminated by an uncorrelated signal or by another correlated light curve, the lag measurement will be affected. We check the effect of dilution with three scenarios shown in Fig.~\ref{dilution}:

\begin{enumerate}
\item Adding an incoherent component: We simulate 1000 pairs of coherent light curves with a 50 second time delay. We contaminated a fraction of one of the coherent light curves with an uncorrelated light curve (different realization of the random Fourier amplitude and phase), and measured the lag and the coherence. We find that the coherence drops, but the value calculated for the lag is still accurately recovered.
\item Mixing coherent components: We simulated 1000 coherent light curves with a 50 second time delay, and mix one light curve with a fraction of the other light curve. We find the lag to be a simple average of the mixed components, while the coherence remains at 1. 
\item Two varying components: We simulate one set of 1000 coherent light curve pairs offset by 50 seconds, and another set of 1000 light curve pairs offset by 20 seconds. The light curves offset by 50 seconds do not have the same realization of the Fourier amplitude and phase as the 20-second set. Thus, the two components are variable, but not coherent. We dilute the 50-second pairs with an increasing amount of the 20-second pairs and find that the recovered lag is a weighted contribution from both components.  This is representative of a scenario in which there are two variable components that vary incoherently with each other, have different lags and different spectral shapes.  This simulation shows that the measured lag can be a weighted contribution from mixed spectral components, and particularly, that low coherence may be an indication of this mixing.
\end{enumerate}

For clarity, Fig.~\ref{dilution} shows the effect of dilution without the addition of Poisson noise into the signal. Even with the addition of Poisson noise (equivalent to that seen in the data), the effects of dilution are the same.

These three scenarios show the signal can be recovered despite contamination by an incoherent signal. Also, if signals are mixed, we will see a proportional drop in the measurement of the lag difference. We also show that requiring a coherence of 1 for a reliable lag measurement is overly conservative. 

In the context of the lag spectra for 1H0707-495, we see from these simulations that in each energy band, the lag is not just due to the reflection or the power law alone, but rather from a contribution of both.  The effect of dilution has been discussed by \citet{zoghbi11}, and in the following section, we take this idea further by modelling the lag spectrum.

\section{Simulating the Lag Spectrum}
\label{lag_model}

In the previous section, we showed the effect of dilution on the lag spectrum. In this section, we show how the lag spectrum can be decomposed via simulations by accounting for dilution in the signal. We offer these simulations as an illustration of a consistent reflection model that follows the data, and not as a statistical fit.  We first review our simple reflection model for the time-integrated spectrum, and then show how the lag-energy spectrum can be modelled in light of a reflection scenario.

\subsection{Modelling the time-integrated spectrum}
\label{spec}

In this section, we briefly describe our model for the time-integrated energy spectrum of the entire 1.3 Ms data set. While the energy spectrum is not the focus of this paper, we will rely on the spectral model for simulating the lag-energy spectrum in the next section.

Because we are working with such a large amount of data, we find that our model fit is completely dictated by the spectrum below 1~keV, where the signal-to-noise is highest.  This is ameliorated when including a 2 per cent systematic uncertainty to all bins, which is a rough, but low-end, estimate of the calibration uncertainty.  This allowed the high energy component (and in particular the clean profile of the Fe K$\alpha$ line) to contribute to the model fit. 
We also split the 1.3 Ms data into 4 flux-resolved spectra, and fit the same models, to understand how the spectra behave in different states. We find no significant difference in the spectral shape aside from changes in the nomalization.  While ionization changes could be expected in different states, so too could we expect density change if the accretion rate changes. Therefore it is perhaps not surprising that the spectral shape remains significantly unchanged between flux states.

We first follow previous spectral work by \citet{fabian12} and \citet{dauser12} in fitting the spectrum with 2 reflection models that are irradiated by a single, steep power law.  The reflection models consist of ionized reflection from an accretion disc \citep[{\sc reflionx} ;][]{ross05}, which is relativistically smeared ({\sc kdblur2f}).  We also include a blackbody to account for both direct disc photons, and thermally reprocessed photons from an irradiated disc. 
This model yields $\chi^{2}/\mathrm{dof} = 1940/1528 = 1.27$. 

\citet{dauser12} showed that adding absorption from a highly ionized outflow \citep[{\sc swind1} ;][]{gierlinski04} significantly improves the fit. In our modelling of the entire data set, we also find this to be a good fit ($\chi^{2}/\mathrm{dof} = 1615/1524 = 1.06$), and agree with the possibility of a highly ionized outflow. However, as this additional component does not significantly change any of the other fit parameters, it does not affect our simulations of the lag-energy spectrum, and we do not consider it further in this paper. 

Given our results from the covariance spectrum discussed in Section \ref{covariance}, we examined a model in which the 2 reflectors are irradiated by 2 power laws with different spectral indices.  The spectral hardening of the covariance spectrum at high frequencies suggests that there could be two coronal regions, one that originates close to the centre and produces a hard power law with short timescale variability, and another more extended region which produces a steeper power law with longer variations. Both power law components likely contribute to the reflection.
We simplify this complex model of two power laws and two reflectors by assuming the same emissivity profile for both reflectors.  We assume a maximally spinning black hole with a steep emissivity that flattens to $q=3$ at a break radius of $2.5~r_{\mathrm{g}}$ (where $r_{\mathrm{g}}=GM/c^2$).  This model results in the best fit of $\chi^{2}/\mathrm{dof} = 1638/1526 = 1.07$.

This picture of an extended corona is congruous with observations from \citet{fabian12}, when 1H0707-495 went into a low state, which was interpreted as emission produced by an irradiating source within 1 $r_{\mathrm{g}}$ of the event horizon.  The photon index for this low state observation was $\Gamma = 2.71^{+0.1}_{-0.07}$. In this study, the hard power law in our time integrated spectrum is $\Gamma \sim 2.65$, and in the covariance spectrum at high frequencies is $\Gamma = 2.89 \pm 0.16$.

\begin{figure}
\includegraphics[width=\columnwidth]{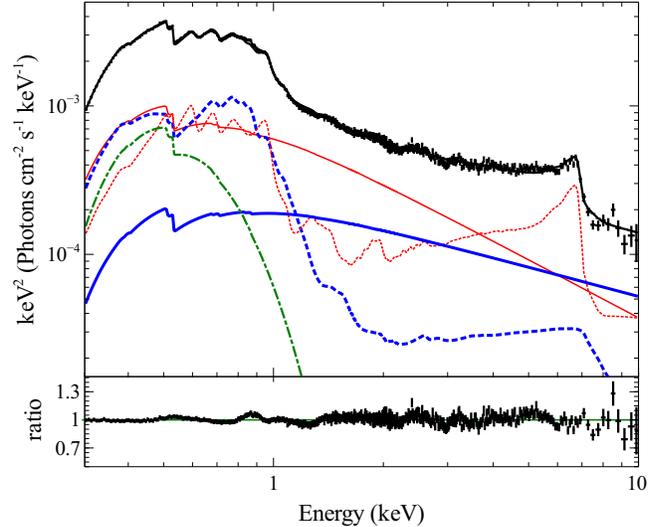}
\caption{Time-integrated energy spectrum for the entire 1.3 Ms exposure. The data are modelled with a blackbody (green dot-dash), 2 power laws (thick blue and thin red solid) and 2 reflection spectra (low ionization reflector, $\xi \sim 50$ irradiated by the soft power law as a thin red dotted line; and high ionization, $\xi \sim 500$, irradiated by the hard power law as the thick blue dotted line). We use the fraction of each component to the total model in order to weight the light curves in our simulations of the lag-energy spectrum. }
\label{energy_spec}
\end{figure}

Fig.~\ref{energy_spec} shows our best-fit model with 2 reflectors and 2 powerlaws. The spectral parameters are shown in Table \ref{spec_param}.  This model will be used in the following section for simulating the lag-energy spectra.

\begin{table}
\centering
\begin{tabular}{l l l}
\hline
{\bf Component} & {\bf Parameter} & {\bf Value}\\
\hline
Galactic absorption& $N_{\mathrm{H}}(10^{22}$ $\mathrm{cm}^{-2})$ $^{\alpha}$& $0.058$ \\
intrinsic absorption& $N_{\mathrm{H}}(10^{22}$ $\mathrm{cm}^{-2})$ & $0.031$ $^{+0.005}_{-0.002}$ \\
\hline
blackbody & $A_{\mathrm{bb}} \times 10^{-5}$ &$ 4.5$ $^{+1.1}_{-0.6}$\\
& $kT_{\mathrm{bb}}$ (eV) & $88.8^{+2.2}_{-3.3}$\\
\hline
power law 1 & $A_{\Gamma_{1}} \times 10^{-4}$ &$ 2.3$ $^{+0.6}_{-1.7}$\\
& $\Gamma_{1}$& $2.66^{+0.04}_{-0.07}$ \\
\hline
power law 2 & $A_{\Gamma_{2}} \times 10^{-4}$ &$ 7.4$ $^{+0.4}_{-1.3}$\\
& $\Gamma_{2}$&$ 3.30^{+0.0}_{-0.02}$ \\
\hline
kdblur2f&$r_{\mathrm{in}} (r_{\mathrm{g}})$& $1.235^{+0.01}_{-0.0}$ \\
&$q_{1}$ & $10.0$ $^{+0.}_{-0.1}$\\
&$r_{\mathrm{out}} (r_{\mathrm{g}}) $ $^{\alpha}$& 400.0 \\
&$i$ ($^{\circ}$) & $46.8$ $\pm 0.4$\\
&$r_{\mathrm{break}} (r_{\mathrm{g}})$ &$2.49^{+0.03}_{-0.02}$ \\
&$q_{2}$ $^{\alpha}$& $3.0$  \\
\hline
reflionx 1 & $A_{\mathrm{refl}_{1}} \times 10^{-6}$ &$ 0.13 \pm 0.02$\\
&$\xi_{\mathrm{refl}_{1}}$ (erg cm s$^{-1}$) & $484$ $^{+9}_{-34}$\\
&$Z_{\mathrm{Fe}}$ & $13.4$ $^{+0.5}_{-0.4}$ \\
\hline
reflionx 2&$A_{\mathrm{refl}_{2}} \times 10^{-6}$ & $43.0$ $^{+2.4}_{-3.3}$\\
&$\xi_{\mathrm{refl}_{2}}$ (erg cm s$^{-1}$) & $49.7$ $^{+0.4}_{-4.0}$\\
\hline
$\chi^{2}/\mathrm{dof}$ & 1638/1526 = 1.07 & \\
\hline
\multicolumn{3}{l}{$^{\alpha}$ frozen}\\
\end{tabular}
\caption[Best-fit Spectral Parameters]{Best-fit spectral parameters for the time-integrated spectrum.}
\label{spec_param}
\end{table}

\subsection{Modelling the lag-energy spectrum}

Similar to the Monte Carlo simulations in Section~\ref{simulations}, we simulate 1000 sets of three identical (but time-shifted) light curves to represent three spectral components: a blackbody, power law, and reflector.  The light curves are simulated without Poisson noise, but as noted in Section~\ref{dilution_sec}, the general effects of dilution on the lag are the same. The signal in each energy band is composed of a fractional contribution from each of these three components, weighted by their contributions in the time-integrated energy spectrum shown in Fig.~\ref{energy_spec}. 

We note that the time-integrated spectrum is not the ideal measure for the weighting because each of the spectral components likely contributes more in certain frequency ranges than in others.  However, we find that we can still model the lag by making some assumptions about the variability of the components in the time-integrated spectrum: 

\begin{enumerate}
\item The blackbody component is likely composed of both direct disc photons (that would be appreciable in the low frequency signal) and thermal photons that are produced by the irradiating continuum heating up the surface of the disc (discernible in the high frequency signal).  Because we cannot distinguish between the low and high frequency contributions in the time-integrated spectrum, we use the entire blackbody when calculating the weighting for both the high and low frequency lags.
\item  We have seen in Section \ref{covariance} that the power law does not contribute the same at all frequencies, and there is likely mixing between at least two power-law components at different frequencies.  For the low frequencies, the fluctuations propagate inwards and are upscattered in the soft power law emitting corona first, and then later in the inner hard corona.  In our simulations of the low frequency lag spectrum, we include the signals from both the hard and soft power-law components. For the high frequencies, where we are presumably probing closer to the inner regions, the hard power-law component dominates, and therefore in our high frequency model, we only simulate the hard power-law component. 
\item We assume that the lag due to reflection is most appreciable on short timescales, and so we only include the reflection component when modelling the high frequency lag spectrum. It is expected that reflection contributes somewhat at low frequencies as well, but the signal is dominated mostly by the variability from propagating viscous fluctuations. 
\end{enumerate}

\begin{figure}
\begin{center}
\includegraphics[width=\columnwidth]{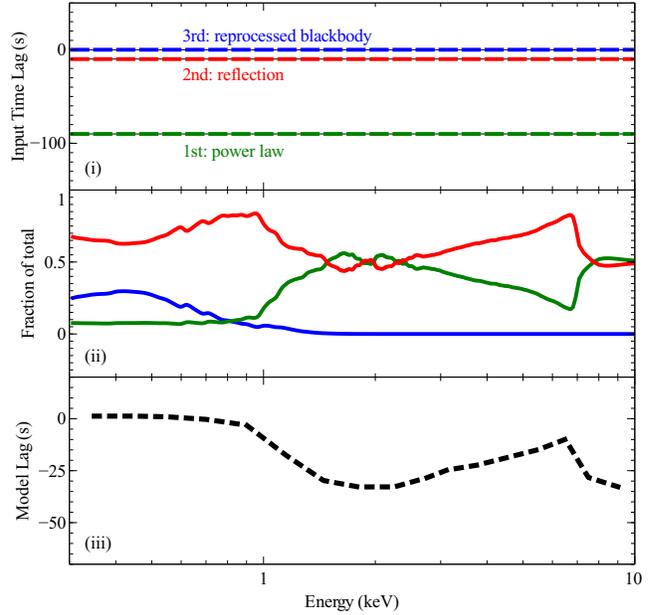}
\caption{Illustration demonstrating how the high frequency lag spectrum is simulated.(i) The input time delays for the hard power law (green), reflection (red) and blackbody (blue) as a function of energy. The power law responds first, followed by the reflection 80 seconds later, and the reprocessed black body 10 seconds after that.  (ii) The fraction of the hard power law, reflection and blackbody to the total time-integrated spectrum. (iii) The simulated lag that is compared to the data in bottom panel of Fig.~\ref{lag_sim}.  Each signal in (i) is weighted by its contribution in (ii), and the total time lag is then summed to make the simulated lag spectrum in (iii).}
\label{explaination}
\end{center}
\end{figure}

The time delay between the contributing spectral components is assumed to be constant for this simple model (i.e. assuming a delta function transfer function).  We input reasonable expectation values for the time delays which are obtained from the geometry implied by our model. Table \ref{tab:time_delay} gives the time delays for both the low and high frequency lag spectral models. We reiterate that the smaller the value of the lag, the earlier it responded.  For instance, at low frequencies, the power law lags behind the blackbody, and therefore its input lag is greater. At high frequencies, the power law leads the reflection, and so its input lag is more negative than the reflection lag. In other words, at high frequencies, the power law responds first, followed by reflection 80 seconds later, and finally followed by the reprocessed blackbody 10 seconds after the reflection. Fig.~\ref{explaination} illustrates how the signals from each component are weighted by the time-integrated 
spectrum to produce the simulated lags.

\begin{table}
\centering
\begin{tabular}{c|c|c}
\hline
 & Low Freq. Lags (s) & High Freq. Lags (s)\\
\hline
blackbody & 0 & 0 \\
$\Gamma_{\mathrm{hard}}$ & 1000 &-90 \\
$\Gamma_{\mathrm{soft}}$ & 400 & -- \\
reflection & -- & -10 \\
\hline
\end{tabular}
\caption{Selected time delays for each component in the high and low frequency lag simulations.  The smaller the value of the lag, the earlier it responds.}
\label{tab:time_delay}
\end{table}

\begin{figure}
\begin{center}
\includegraphics[width=\columnwidth]{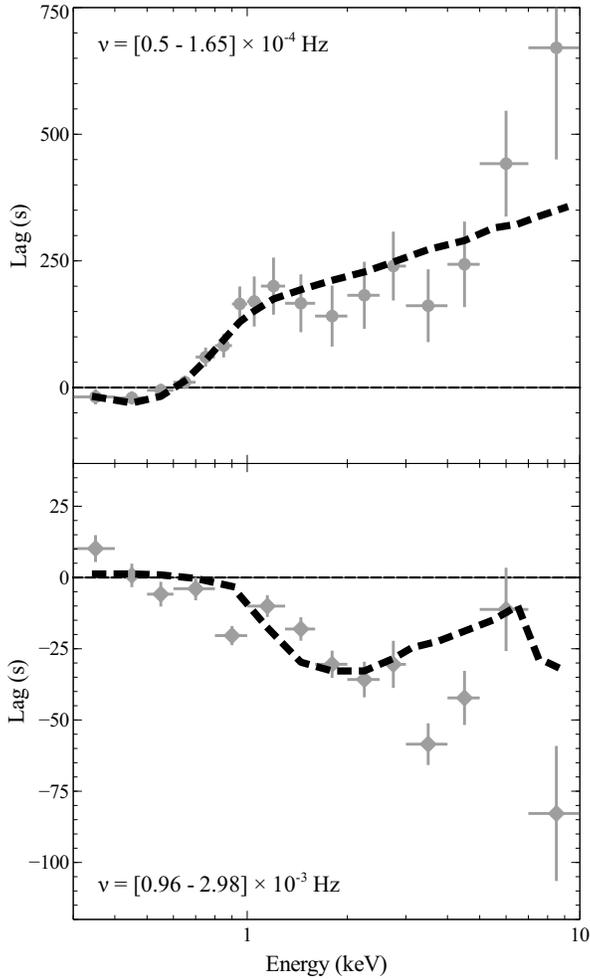}
\caption{Monte Carlo simulations of the low and high frequency lag-energy spectrum (black dashed lines) where the weighting of the signal from the blackbody, power law and reflection components are determined by their fractional contribution to the time-integrated energy spectrum.  Time delays between all components are constant.}
\label{lag_sim}
\end{center}
\end{figure}

The time lag is then calculated in the usual way, with respect to the reference signal from 0.3 -- 0.8~keV.  The top panel in Fig.~\ref{lag_sim} shows the simulated lag spectrum in the frequency range $\nu = [0.5 - 1.65] \times 10^{-4}$~Hz. Notice how the step up from 0 seconds to $\sim$~200 seconds arises naturally due to the time delays between the thermal emission in the disc and the continuum emission in the corona. This can be easily explained by the fractional contribution of the blackbody and two power-law components in each band compared to the reference.  For instance, at 3~keV, the signal is roughly composed of two parts soft power law (with lag $= 400$~s) and one part hard power law (lag $= 1000$~s), amounting to an average lag of $\sim 600$ seconds.  This lag is compared to the reference band, which is roughly composed of 30\% blackbody (lag $= 0$ s), 10\% hard power law and 50\% soft power law, amounting to an average lag of $\sim 300$~s. The lag at 3~keV, is therefore calculated to be $\sim 300$~s.

\begin{figure}
\begin{center}
\includegraphics[width=\columnwidth]{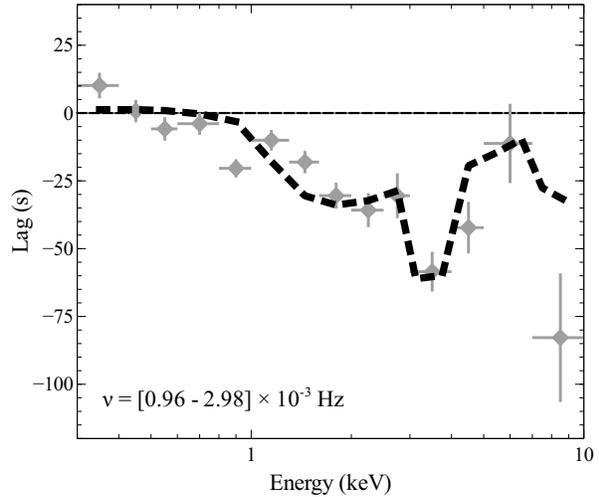}
\caption{Monte Carlo simulations for the high frequency lag spectrum.  This figure has the same model as the bottom panel of Fig.~\ref{lag_sim}, however, we now account for a faster response from the red wing of the Fe K$\alpha$ line. 
In our model, the red wing of the line responds 15 seconds after the continuum emission, while the bulk of the reflected emission responds 80 seconds after the continuum.}
\label{redwing}
\end{center}
\end{figure}

The bottom panel of Fig.~\ref{lag_sim} shows the lag in the frequency range $\nu = [0.96 - 2.98] \times 10^{-3}$~Hz.  The model follows the general shape of the data, and suggests that reflection is playing a major role in diluting the intrinsic lag of the power law. However, the model is particularly unsatisfactory in the 3--4~keV bin. 
At 3--4~keV, the red wing of the iron K$\alpha$ line dominates the reflection spectrum. This is the reflected emission from the innermost regions of the disc, and therefore, it is the signal that responds first. To account for this, we input a time lag of -75~s for the reflection in this bin, which amounts to the reflection lagging behind the power law by just 15 seconds.  Therefore, the effect of dilution on the power law due to reflection is less in this band, making the lag appear to be lower. Fig.~\ref{redwing} shows the results of the simulations with the addition of a fast response from the red wing of the 
Fe K$\alpha$ line.  The addition of the response from the red wing significantly improves the fit to the data.

\section{Discussion}

\subsection{The Low Frequency Lags}

The large positive lag at low frequencies has been explained as the viscous propagation of accretion rate fluctuations in the disc \citep{arevalo06,wilkinson09,uttley11}.
If the energy released from accretion is dissipated in the corona (e.g. through magnetic field lines), fluctuations in the disc could cause correlated variability in the corona.
Our analysis of the hard lag spectrum is consistent with this model.  We find zero lags with respect to the reference band (0.3 -- 0.8~keV) at energies below 1~keV (where the blackbody is appreciable), and above 1~keV, where the power law begins to dominate, we compute a lag behind the reference of 200~s up to 600~s. Considering the effect of dilution in our simulations, we find that the intrinsic time delays are likely to be greater ($\sim 400$~s to $\sim 1000$~s).    

Furthermore, our frequency-resolved analysis of the low frequency lag-energy spectrum (top panel, Fig.~\ref{lag_spec}), shows that for variations on longer timescales (`very-low' frequencies, $\nu = [0.5 - 1.65] \times 10^{-4}$~Hz), the blackbody component contributes less. These very-low frequency variations presumably originate from further regions from the centre, where the temperature of the disc is cooler, and may be out of the spectral sensitivity of {\em XMM-Newton}.  At higher frequencies, ($\nu = [1.65 - 5.44] \times 10^{-4}$~Hz), the variations originate closer in, where the disc is hotter, and therefore, the contribution from the blackbody is greater.  Our simulations also confirm that if we increase the fractional contribution of the blackbody (as we expect for smaller radii in a Shakura--Sunyaev disc), we describe the general shape of the lag-energy spectrum for the `low' frequencies in the top panel Fig.~\ref{lag_spec}.

The low frequency lag spectrum shows a steady increase above 1 keV.  This upward trend has been observed in stellar mass black holes, such as Cyg X-1 \citep{nowak99}. \citet{kotov01} argued that these lags could be due to radial dependence of the spectral index, in which it is hardest in the innermost regions, and becomes softer at further distances.  Perturbations in the accretion flow that are introduced at different radii (and therefore different timescales) propagate inward on diffusion timescales, which cause the observed power law shape of the lag spectrum.  This model for Galactic black holes can also explain the upward trend seen in the lag spectrum of 1H0707-495. The data are well described by our simulations, which follow the Kotov model of the soft power law responding first followed by the hard power law from the innermost regions. 

Finally, this interpretation of an extended corona is consistent with relativistic models by \citet{wilkins11} that show that a majority of the power law comes from a small region within 5 $r_{\mathrm{g}}$, while the rest comes from further out.

\subsection{The High Frequency Lags}

The negative lags at high frequencies have been explained as the light travel time between continuum emission produced in the corona and reflected emission from the inner regions of the disc.  The soft lag spectrum in bottom panel of Fig.~\ref{lag_spec}b is generally consistent with this model in that variations at energies dominated by the reflection spectrum are temporally coincident with those of the reference band, while the energies dominated by the power-law component lead those variations.  This result is consistent with previous results, and now with the new long dataset, we are able to define the lag spectrum much more clearly, revealing structure at the Fe K$\alpha$ line.  

An important result from the simulations in Fig.~\ref{lag_sim} is to show that the actual time delays between the power law emission and the reflection are not necessarily the same as the lags we measure because of the effect of dilution. 
We find that for much of the reflection, the actual light travel lag is more on the order of 80 seconds, corresponding to a coronal height of $\sim~4$ $r_{\mathrm{g}}$ (assuming a black hole mass $\sim~2 \times 10^{6} \mathrm{M}_{\odot}$). At 3 -- 4~keV, the red wing of the iron K$\alpha$ line, we find that the light travel time is $\sim~15$ seconds, corresponding to emission from within 1 $r_{\mathrm{g}}$ of the event horizon.
Future work using realistic transfer functions, which take into account the spatial extent of the source and the inclination of the disc, will improve our soft lag spectral models, and give us a better understanding of the geometry of the innermost regions closest to the black hole.

\section{Conclusions}

We have presented the timing analysis of 1H0707-495 using 1.3~Ms of data taken over 10 years with {\em XMM-Newton}.  
We have simulated the lag spectra using a model that accounts for dilution from different spectral components in calculating the lag.  This simple model for the hard and soft lag spectra shows complicated features that can be explained by reflection.  We have also shown through an analysis of the covariance spectrum and the time-integrated spectrum, that there is evidence for a more complicated continuum emitting region, which produces a harder power law closer to the centre, and a softer power law at larger radii.  Further work needs to be done to understand the complexities of the power law emitting region.  This, and future work with more realistic transfer functions for the response of the accretion disc will be beneficial in uncovering the precise origin of the X-ray emission in the local environments of supermassive black holes.

\section*{Acknowledgments}

This work is based on observations obtained with {\em XMM-Newton}, an ESA science mission with instruments and contributions directly funded by ESA Member States and NASA.  EK is supported by the Gates Cambridge Scholarship.  EK thanks Dom Walton for useful discussions.  ACF thanks the Royal Society.  The authors thank an anonymous reviewer for helpful comments.

\appendix

\renewcommand{\thefigure}{\Alph{figure}}

\section*{Appendix : Checking the Background for Lags}

\begin{figure}
\includegraphics[width=\columnwidth]{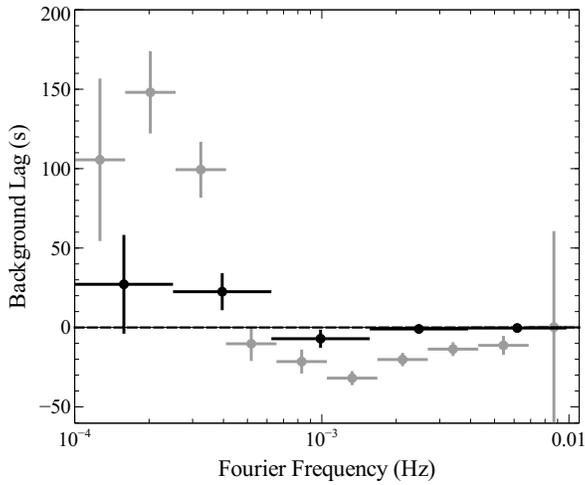}
\caption{Background lag-frequency spectrum in black, with source lag-frequency from Fig.~\ref{lag_freq} in grey for reference.  While the general shape of the source and background lag-frequency spectra are similar, the background lags are not significant.  We perform this background analysis to confirm that the lags in Fig.~\ref{lag_freq} are intrinsic to the source.}
\label{bkg}
\end{figure}

To ensure that the lags in Fig.~\ref{lag_freq} are intrinsic to the source, we replicated the analysis on background light curves, and found no significant lag.  
The same Fourier analysis, as described in the previous section, was performed using the hard and soft background light curves of the 14 observations. The results are shown in Fig.~\ref{bkg}. The general shape of the background lag-frequency spectrum is similar to that of the source, but the amplitude of the lags is significantly smaller.  This could perhaps be caused by contamination of the background by source photons.  A complete study of these mild background fluctuations is beyond the scope of this paper, but we do conclude that the lags seen in Fig.~\ref{lag_freq} are intrinsic to the source.

\label{lastpage}

\end{document}